\documentclass[%
 reprint,showkeys,
nofootinbib,
 amsmath,amssymb,
 aps,
]{revtex4-2}
\usepackage{graphicx}% Include figure files
\usepackage{dcolumn}% Align table columns on decimal point
\usepackage{bm}% bold math
\usepackage{hyperref}% add hypertext capabilities
\usepackage{derivative}

\usepackage{booktabs}%\usepackage[showframe,%Uncomment any one of the following lines to test 
\usepackage{paralist}
\newcommand{\ie}{\textit{i.e.}}
\newcommand{\eref}[1]{eq.~\ref{#1}}
\newcommand{\fref}[1]{fig.~\ref{#1}}

\begin{document}

\preprint{APS/123-QED}

\title{Cooperation in a non-Ergodic World on a Network - Insurance and Beyond}% Force line breaks with \\

\author{Tobias Wand}
\email{Corresponding author's e-mail address:\\ t\_wand01@uni-muenster.de}
\affiliation{Institute of Theoretical Physics, University of Münster}
\affiliation{Center for Nonlinear Science Münster, University of Münster}

\author{Oliver Kamps}
\affiliation{Center for Nonlinear Science Münster, University of Münster}

\author{Benjamin Skjold}
\affiliation{London Mathematical Laboratory}
\affiliation{Danish Research Centre for Magnetic Resonance}

\date{\today}% It is always \today, today,

\begin{abstract} 
Cooperation between individuals is emergent in all parts of society, yet mechanistic reasons for this emergence is ill understood in the literature. A specific example of this is insurance. Recent work has, though, shown that assuming the risk individuals face is proportional to their wealth and optimising the time average growth rate rather than the ensemble average results in a non-zero-sum game, where both parties benefit from cooperation through insurance contracts. In a recent paper, Peters and Skjold present a simple agent-based model and show how, over time, agents that enter into such cooperatives outperform agents that do not. Here, we extend this work by restricting the possible connections between agents via a lattice network.
%Cooperation can emerge endogenously between individual agents if they are subject to non-ergodic dynamics and this phenomenon occurs even if all agents are only optimising their own long-term rewards. In this article, we simulate agents and their wealth dynamics under repeated gambles on a lattice network and allow next neighbours to cooperate by signing an insurance contract. 
Under these restrictions, we still find that all agents profit from cooperating through insurance. We, though, further find that clusters of poor and rich agents emerge endogenously on the two-dimensional map and that wealth inequalities persist for a long duration, consistent with the phenomenon known as the poverty trap. 
%While the appearance of rich and poor neighbourhoods superficially resembles a form of kin selection, it emerges endogenously from the model dynamics. 
By tuning the parameters which control the risk levels, we simulate both highly advantageous and extremely risky gambles and show that despite the qualitative shift in the type of risk, the findings are consistent.
%Parameter variations show that our model then transitions between two forms of cooperation: Insuring the detrimental risk of another agent or offering another agent a deterministic fixed income in return for its fluctuating, but positive output.
\end{abstract}

 %Use showkeys class option if keyword
\maketitle

%\tableofcontents

\section{Introduction}
Self-organisation of individuals is observed in many systems and is a cornerstone of complexity sciences~\cite{Zhang1991,selforganisation_economic,strogatz_book,selforganisation_economics_irreversibility_entropy}. Mutual cooperation between agents exists in many forms, one such being insurance. The insurance industry is one of the largest industries in the modern economy and has existed for centuries. Yet, we find that mainstream economics, which is rooted in expected value theory, does not provide a satisfactory reason for its emergence. As an alternative approach, Peters and Adamou argue in \cite{peters2017insurance} that the system in which insurance takes place is more reminiscent of a multiplicative system (i.e. that the risk is proportional to the wealth) and note that the increments in such a system are not ergodic. Using time averages rather than ensemble averages, they show that there exist regimes in which insurance contracts are favourable for both buyer and seller, which calls for such contracts to be made. In a recent paper, Peters and Skjold~\cite{peters_2023_editorial,ole_peters_2023_7994190} explore this idea using a simple agent-based model, which shows that agents who cooperate via insurance contracts over time systematically outperform agents who do not.\\
Inspired by the 2d Ising model on a lattice and its many applications to socio-economic phenomena ~\cite{Ising1925,Onsager_Ising_PhysRev.65.117,Nowak1992,Ising-like_SZNAJD-WERON,Stauffer2008,Sornette2014}, we expand the model proposed in \cite{peters_2023_editorial,ole_peters_2023_7994190} to large ensembles on a lattice network. Now, each agent is placed in a neighbourhood of other agents and can only cooperate with nearby agents, giving rise to the formation of spatial patterns in our model similar to \cite{SegregationModel,Nowak1992}. \\

In this paper, we first present a brief background of non-ergodicity in economics and illustrate the implication of broken ergodicity with a simple coin toss example before detailing the insurance paradox and its treatment in the context of ergodicity economics in section~\ref{sc:IntroductionEE}. Next, we describe the setup of our implementation of the agent-based insurance model on a network and the mathematical techniques used to analyse its results in section~\ref{sc:MethodsandModels}. Results for a specific parametric setting are presented in section \ref{sc:TypicalResults} before we perform a parameter scan by varying the parameters $c$ and $r$ representing the relative costs or rewards of the risk in section~\ref{sc:Parameterscan}, which reveals different clustering regimes in our model. Finally, we conclude with a summary and discussion of our work in section~\ref{sc:Conclusion}.

\section{Introduction to Ergodicity Economics}
\label{sc:IntroductionEE}
In an ergodic system, an individual trajectory at different time steps $x_i(t_1),\dots,x_i(t_n)$ has the same statistical properties as an ensemble observed at a single point in time $x_1(t),\dots,x_n(t)$. This property makes many calculations easier, and thus the ergodic hypothesis is often explicitly or implicitly assumed for many statistical methods. The core of ergodicity economics is a careful analysis of whether the ergodic hypothesis is true and thus makes an explicit distinction between ensemble averages $\langle x \rangle = \frac{1}{N}\sum_i x_i(t)$ and time averages $\Bar{x} = \frac{1}{T} \sum_t x_i(t)$  for e.g. financial time series~\cite{Script}. Assuming ergodicity implies that $\langle x\rangle = \Bar{x}$ in the limit of large $T$ and $N$. However, it is the exception rather than the norm that this is justified in real-world systems. A simple example of a non-ergodic system is the reinvesting coin toss gamble taken from \cite{Peters_Gell-Mann}.

\subsection{Reinvesting Coin Tosses}
Consider some initial wealth $x(0)$. You are now offered a gamble on which a fair coin is tossed, i.e. with probability $1/2$, you win and with probability $1/2$, you lose. If you win, your wealth is multiplied by a factor of $\alpha_w = 1.5$, and if you lose, your wealth is multiplied by a factor $\alpha_l = 0.6$. We will repeatedly toss the coin and sequentially update your wealth according to the outcome of the toss. Is this a favourable bet? The expected value computed via the ensemble average of identical agents with initial wealth $x$ shows that
\begin{equation}
    \label{eq:CoinTossEnsembleAverage}
    \langle x(t+1)|x(t) = x \rangle = \frac{\left( \alpha_wx + \alpha_lx\right)}{2} = 1.05 x > x(t),
\end{equation}
and therefore implies that the gamble is favourable. However, considering that for a single agent with initial wealth $x(0)$ and enough repetitions $T$, the law of large numbers guarantees that both coin toss outcomes will appear equally often, and we can compute the time average
\begin{equation}
    \label{eq:CoinTossTimeAverage}
     \Bar{x}(T) = \alpha_w^{(T/2)}\alpha_l^{(T/2)} x = 0.9^{(T/2)}x \overset{T\rightarrow \infty }{\longrightarrow} 0.
\end{equation}
Hence, the agent's wealth approaches $0$ almost surely, implying that the gamble is not favourable. As proven via numerical simulations in, for example, \cite{Script}, this seemingly paradoxical situation can be explained by noting that the increasing ensemble average of the wealth trajectory is dominated by the asymmetry of the final wealth. While almost all agents have negligible wealth, few have exponentially increasing wealth, which dominates the ensemble average. 

\subsection{Insurance Paradox}
Consider a very fundamental and simple type of insurance, namely an agent, $A$, faces a risk and another agent, $B$, offers to take over that risk in exchange for receiving a fee, $F$, typically known as the insurance premium. The classical economic analysis of the insurance problem is based on the expected value model, which posits that agent $A$ should accept to pay a fee lower than the expected value of the risk, whereas $B$ should demand a fee higher than the expected value of the risk. This anti-symmetry results in no fee where both agents simultaneously are satisfied, and thus, no contracts should be agreed upon. The orthodox solution is usually either that the agents must have an information asymmetry or that ``risk aversion'' makes the agents deviate from the supposedly optimal decision~\cite{AsymmetricInformation_RiskAversion,arrow1971theory}. We find this unsatisfactory as the former is unjustified ad hoc reasoning and the latter simply raises a new question as to why such risk aversion exits.\\ 
Assuming that the risk an agent faces is proportional to their wealth and optimising time averages, however, Peters and Adamou~\cite{peters2017insurance} show that there exist many situations in which agent $A$ is willing to pay a higher fee than agent $B$ demands. This framework provides a mechanistic reasoning for the existence of insurance contracts and shows that insurance is theoretically advantageous in the long run. These arguments are discussed in more detail in \cite{peters2017insurance,peters_2023_editorial}, and a broader scope of cooperation in non-ergodic systems can be found in \cite{Peters2022_cooperation}. We highlight that behavioural experiments show evidence that humans are capable of heuristically optimising the time average growth rate when faced with risky decisions in laboratory settings \cite{CopenhagenExperiment,Vanhoyweghen2023,Skjold2023}, which gives credibility to the time solution of the insurance puzzle.

\section{Methods and Models}
\label{sc:MethodsandModels}
\subsection{Insurance among Agents}
As we expand on the model in \cite{peters_2023_editorial,InsuranceModelBlogPost,ole_peters_2023_7994190}, it is prudent to review its results alongside its model specifications. This model consists of $n$ agents $i=1,\dots,n$, which are observed at time $t \in (0,T)$. Each agent's wealth is initialised at $1$, $w_i(0) = 1$ $\forall i$, and in each time step, an agent $A$ with wealth $w_A(t)$ is randomly chosen to face a risky gamble: with probability $p$, $A$ either loses a relative amount of its wealth $c\cdot w_A(t)$ or with probability $1-p$ wins a relative amount $r\cdot w_A(t)$. Note that the simulations in \cite{peters_2023_editorial,ole_peters_2023_7994190} use $c=0.95$, $r=0$ and $p=0.05$ as parameters, but that the qualitative results do not depend on the exact parameter choice, as exemplified in \cite{InsuranceModelBlogPost}.\\ 
When faced with a risky gamble, agent $A$ approaches another agent ($B$) for an insurance contract. Agent $B$ will have a minimum fee, $F_{\min}$, where they will accept to take over $A$'s risk and agent $A$ will have a maximum fee, $F_{\max}$, they will accept to pay. If $F_{\min}<F_{\max}$, there exist a range of fees in which signing the contract is mutually beneficial for the agents, such that the risk and the fee is transferred from agent $A$ to agent $B$.\footnote{For stability reasons we restrict contracts to be formed by enforcing $w_B(t)\geq cw_A(t)$ to explicitly ensure agent $B$ can pay for the cost $cw_A(t)$ without defaulting. However, we note that this is not strictly necessary but has no qualitative effect on the results.} To ensure the results are not an conflated by the number of agents used, we distinguish between a time step, $t$, and a time unit, $t_u$, with a time unit being $N^2$ time steps, i.e. each agent is on average chosen to face a risk ones per time unit.\\
The ergodicity economics solution to the insurance problem is to focus on time average growth rates. Recognising that the underlying process of the system is mainly multiplicative\footnote{We recognise that the multiplicative structure of the simulations is not entirely realistic, and even within the simulation, the wealth increments are not fully multiplicative. However, these subtle distinctions are outside the scope of this analysis and part of ongoing research.}, it follows that additive wealth increments are not ergodic, but logarithmic wealth increments are and, therefore, better indicate what happens over time and allow us to assume an underlying growth rate model $g(\cdot)$~\cite{Peters_Gell-Mann,peters2017insurance}. Comparing agents' growth rates with $g_{A,B}^{with}$ and without $g_{A,B}^{without}$  signing an insurance contract at some fee $F$, we can calculate the maximum fee $F_{\max}$ the agent facing the risk (agent $A$) is willing to pay, as well as the minimum fee $F_{\min}$ another agent (agent $B$) is willing to accept. We therefore get that 
\begin{widetext}
\begin{eqnarray}
\label{eq:Fmax}
    & \underbrace{\frac{1}{\Delta t}\ln\left(\frac{w_{A}-F}{w_{A}}\right)}_{g_{A}^{\text{with}}}= \underbrace{\frac{1}{\Delta t}\left( p \ln\left(\frac{w_{A}-cw_A}{w_{A}}\right) 
 + (1-p)  \ln\left(\frac{w_{A}+rw_A}{w_{A}}\right) \right)}_{g_{A}^{\text{without}}} \\ \nonumber
   \Leftrightarrow \quad &F_\text{max}  =w_{A}\left[1-\exp(g_{A}^{\text{without}})\right] 
\end{eqnarray}

for agent $A$ and

\begin{align}
\label{eq:Fmin}
& \underbrace{(1-p)\frac{1}{\Delta t}\ln\left(\frac{w_{B}+F+rw_A}{w_{B}}\right)+p \frac{1}{\Delta t}\ln\left(\frac{w_{B}+F-cw_A}{w_{B}}\right)}_{g_{B}^{\text{with}}} = \underbrace{0}_{g_{B}^{\text{without}}}
\end{align}
\end{widetext}
for agent B, where $g_B^{\text{without}} = 0$ because without the insurance, only agent $A$ experiences any change during the observed time step. Eq.~\ref{eq:Fmin} does not have a general closed-form solution but does have for $p=1/2$. We, therefore, use this parameter value for our simulation.\footnote{This is especially useful because numerical optimisation of \eref{eq:Fmin} breaks down for sufficiently small wealth levels.}

For $p=1/2$, we get that $F_{\max}$ and $F_{\min}$ are given by
\begin{align}
\label{eq:Fmax_Fmin}
     F_{\max} &= w_A - (w_A + rw_A)^{0.5}(w_A-cw_A)^{0.5} \\ \nonumber
     F_{\min} &= -w_B  + \frac{1}{2} \sqrt{4w_B^2 + (c+r)^2w_A^2 } + \frac{(c-r)w_A}{2}.
\end{align}

For suitable $w_A$ and $w_B$, $F_{\min} <F_{\max}$ is fulfilled and therefore, an interval of fees $F$ in $[F_{\min} , F_{\max}]$ exist for which insurance contracts are mutually beneficial for both agents. For simplicity, choose then the midpoint fee $F = \frac{1}{2}(F_{\min} +F_{\max})$ to sign the contract. As shown in Fig. 3 of \cite{peters_2023_editorial}, the agents who sign contracts based on this criterion outperform agents who do not (for example, expected value optimisers). Interestingly, the simulations show how ``large'' agents, who act as the insurance company for all other agents, emerge. However, if such an agent becomes ``too large,'' they lose the ability for other agents to insure its large risk and thus, its growth rate declines until it is no longer larger than the other agents.

\subsection{Simulation on a Network}
The model in \cite{peters_2023_editorial} is deliberately simple but showcases two very interesting findings: First, all agents participating in the insurance contracts profit compared to the uninsured agents. Second, even though the cooperation reduces the overall wealth inequality compared to a model without insurance, it still endogenously creates large wealth differences, leading to the emergence of a large insurance-company-like agent.\\
However, the well-mixed setup in which all agents can approach each other is only realistic for small systems. Therefore, we propose to perform the simulation on a lattice and allow each agent to approach only its nearest neighbours to negotiate an insurance contract. The lattice simulation allows us to investigate the spatial clustering of agents based on relative wealth levels. We use a lattice coordination number of four and periodic boundary conditions such that the first and last agents of each row and column are treated as neighbours. Throughout the simulations, we use a square lattice with length $N=64$, i.e. $N\times N = 2^{12}$ agents~\cite{Code}. %Also, we provide a brief discussion of the model behaviour when allowing for interactions beyond the nearest neighbours in the conclusion. 
\subsection{Clustering Evaluation Methods}
The main difference between our model and the one presented in \cite{peters_2023_editorial} is the spatial constraints, which give rise to the analysis of spatial clustering. We, however, first investigate the temporal clustering and compare that to the temporal clustering from the model in \cite{peters_2023_editorial}.

%We now present methods to evaluate the spatial clustering into neighbourhoods and the temporal clustering into persistently rich and poor agents. These methods will be used to formalise the observed results from section \ref{sc:TypicalResults} with mathematical quantities.

\subsubsection{Temporal Clustering}
To evaluate the temporal clustering of agents, \ie whether the rich stay rich and the poor stay poor, we use Spearman's rank correlation. All agents are ordered in an enumeration $i=1,\dots,N^2$, and we record their ranks $\rho_i(t_u)$ as the $\text{k}^\text{th}$ richest agent for each time unit $t_u$. %Note that $\rho(t)\in\mathbb{N}^{N^2}$ is a permutation of $(1,\dots,N^2)$. 
Spearman's rank correlation is now the regular Pearson's correlation between $\rho(t_u)$ and $\rho(t_u^\prime)$. For each time lag $\Delta t_u$, multiple pairs $t_u-t_u^\prime = \Delta t_u$ exist, and therefore, we calculate Spearman's correlation for all pairs with a given time lag and compute the mean and standard deviations for all of them.

\subsubsection{Spatial Clustering}
%\subsubsection*{Clustering Coefficient}
A naive approach to evaluate the neighbourhood formation of the richest or poorest quantile is formulated by defining a clustering coefficient $\kappa_{clust}$ for each group as the relative amount of agents in this group which have at least one neighbour from their own group. For the definition of the groups via deciles and for a sufficiently large ensemble $N\gg 1$,\footnote{With sufficiently large, we mean that when checking whether any of the four sites is occupied by another member of the same decile, it can be modelled as sampling without replacement from a group with a relative size of $\frac{1}{10}$. This holds if $\frac{0.1\cdot N^2 - 1}{N^2} \approx \frac{1}{10}$.} the expected clustering coefficient for a purely random ensemble can be calculated as 
\begin{equation}
    \label{eq:ExpectedClustering}
    \kappa_{clust}^{random} = 1 - (1-0.1)^4 \approx 0.34,
\end{equation}
and is the chance that none of the i.i.d. four neighbourhood sites is occupied by another member from the in-group. Approximate confidence intervals of this quantity can be calculated empirically by simulating a large number of random matrices and evaluating their $\kappa_{clust}$.\\

%\subsubsection*{Spatial Autocorrelation}
An alternative way to evaluate the spatial clustering closer to the framework of Ising-like models is to compute the spatial autocorrelation. To this end, the ensemble matrix is transformed into a binary matrix \textbf{B} with $B_{ij}=1$ if a member of the group under consideration (i.e. the richest or poorest decile) is in the $\text{i}^\text{th}$ row of the $\text{j}^\text{th}$  column and $0$ otherwise. The autocorrelation matrix is then computed via the function \textit{scipy.signal.correlate2d}~\cite{2020SciPy-NMeth} by correlating \textbf{B} with itself and using the settings \textit{mode ="full", boundary = "wrap", fillvalue=0}. This method computes the full discrete linear cross-correlation of \textbf{B} with itself as a 2-dimensional array. 

%\subsubsection*{Spearman's Rank Correlation}
\section{Results: Typical Observations}
\label{sc:TypicalResults}
In an initial run, we set $T_u = 200$, $c = 0.6$ and $r = 1.4$. These parameters are deliberately chosen, such that for an agent without any insurance, this yields the interesting regime of 
\begin{align}
    \langle w(t+1)\rangle = 1.4w(t) > w(t) \\
    \Bar{w}(t+1) = \sqrt{0.96}w(t) < w(t),
\end{align}
i.e. from the ensemble perspective, it is advantageous to keep the risk, but from the time perspective, it is not. This also ensures that the trajectories do not decay too quickly to zero because $\sqrt{0.96} \lesssim 1$, thereby guaranteeing numerical stability of the simulation. %Note that $(c,r,T)$ will be varied to explore other regimes of model behaviour in section \ref{sc:Parameterscan}, but the given values for $(c,r,T)$ will be used here to motivate the use of the evaluation methods and to display interesting and typical model behaviour. 

\subsection{Temporal Clustering}
\begin{figure}[h]
    \centering
    \includegraphics[width = \linewidth]{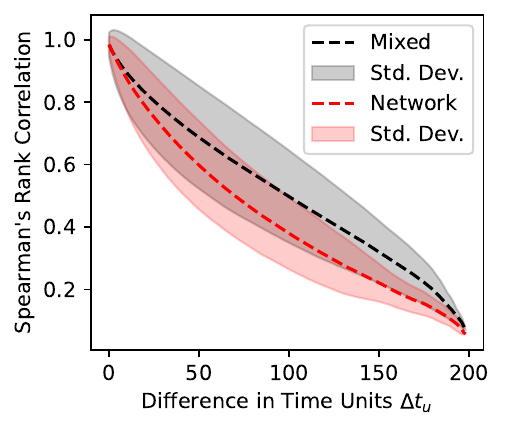}
    \caption{Spearman's rank correlation shows a highly significant correlation in the ranking order of the agents' wealth levels even after long time lags for both our network model and the mixed model in \cite{peters_2023_editorial}. The parameters used for this simulation are $T_u=200$, $N=64$, $c=0.6$ and $r=1.4$.}
    \label{fig:Spearman}
\end{figure}

In \fref{fig:Spearman}, we show Spearman's rank correlation, which reveals that even for large time differences, there is a great ranking correlation showcasing that there is a large memory in the agents' wealth ranking. This means that the general wealth ranking is preserved for extended periods of time, i.e. that, for example, ``large'' agents stay as the ``large'' agents for long. This is similar to what is observed in \cite{peters_2023_editorial} and can be related to the well-documented poverty trap observed in real societies \cite{PovertyTrap_Ghatak2015,Persistence_inequality}. 
%Comparison between the two models shows that the Spearman correlation is slightly lower for the network model than for the well-mixed system, though that their standard deviations overlap to a large degree. Though the network constraints are not necessary to produce the long-term wealth hierarchy shown in \fref{fig:Spearman}, this emergent property has not yet been discussed in the work \cite{peters_2023_editorial,InsuranceModelBlogPost,ole_peters_2023_7994190} and is replicated by the model in our article. For the same parameters $(c,r)$, simulations of the network model show that the wealthiest agent is much more frequently replaced by another, new wealthiest agents when compared to the well-mixed model: While the well-mixed model allows the richest agent to offer contracts to all other agents, the network model prevents this and instead creates local niches where other rich agents can thrive without competing against the market leader.

\begin{figure*}
    \centering
    \includegraphics[width = \textwidth]{ 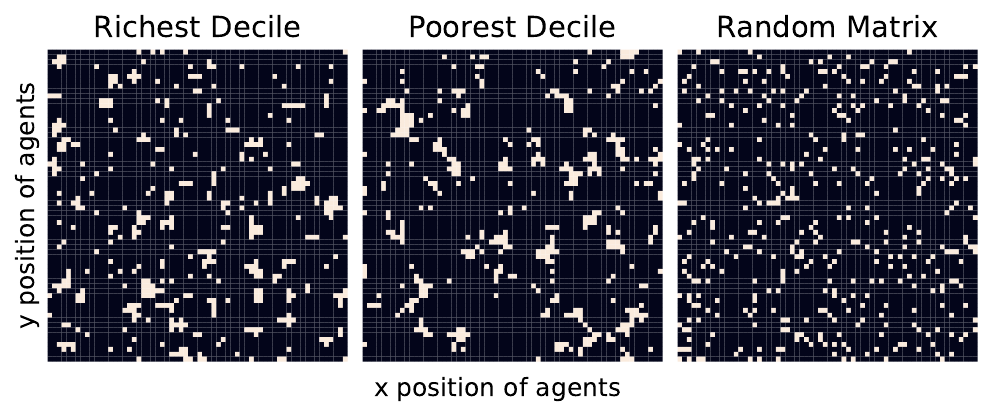}
    \caption{On the $N\times N$ lattice grid, we highlight the richest 10\% of agents (left panel) and the poorest 10\% of agents (middle panel) in our model. As a comparison, we also show a purely random ensemble of agents (right panel) and highlight its 10\% richest agents. Our model shows a marked pattern of clustering into neighbourhoods when compared to the random matrix. The parameters used for this simulation were $T_u=200$, $N=64$, $c=0.6$ and $r=1.4$. }
    \label{fig:Clustering_Rich_Poor_Random}
\end{figure*}

\begin{figure*}
    \centering
    \includegraphics[width = \textwidth]{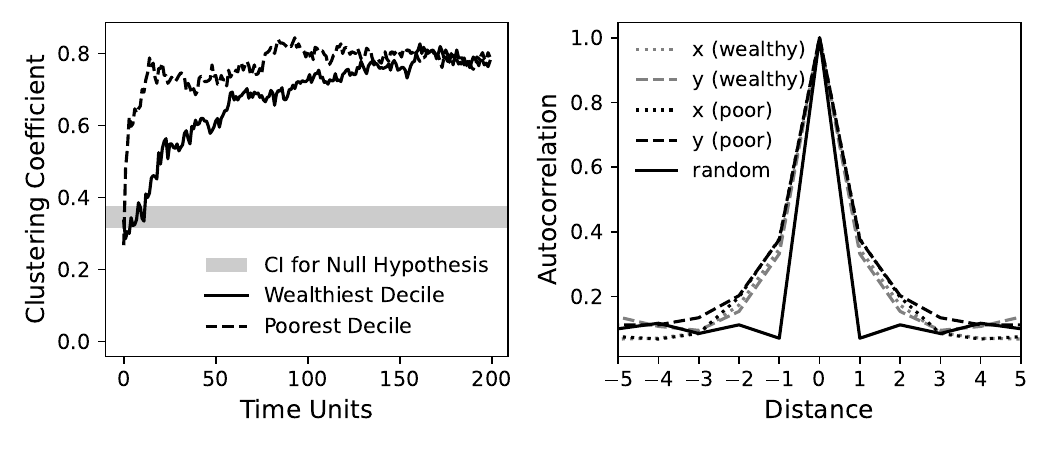}
\caption{\textbf{Left}: Clustering coefficients of the richest and poorest deciles are far above the expectation for random ensembles. While the clustering emerges faster for the poorest than for the wealthiest decile, both seem to converge to similar levels. The confidence interval for the null hypothesis that the geographical distribution is purely random is calculated via eq. \eqref{eq:ExpectedClustering} and sampling random matrices. \textbf{Right}: Autocorrelation function (ACF) for the wealthiest and poorest deciles in x and y direction. All show essentially the same behaviour and a notably higher value at the first nontrivial distance than for the random matrix. The parameters used for this simulation are $T_u=200$, $N=64$, $c=0.6$ and $r=1.4$ for both graphs.}
    \label{fig:Clustering}
\end{figure*}

\subsection{Spatial Clustering}

\begin{figure*}
    \centering
    \includegraphics[width = 0.95\textwidth ]{ 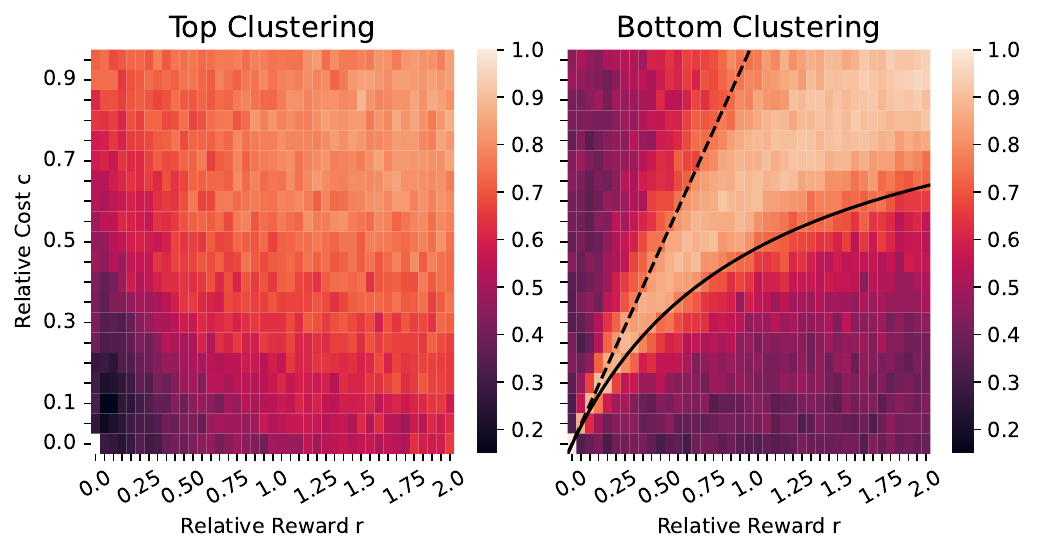}
    \caption{For $T_u=200$ and $N=64$, $c$ and $r$ are varied and the clustering coefficients for the top and bottom decile are recorded after the last iteration. Note that $c=r=0.0$ means the absence of gambling dynamics and, therefore, shows no dynamical behaviour at all. The regime for high clustering of the bottom decile has contours given by the black lines. These lines correspond to the onset of degenerate regimes for $F_{\min/\max}$ and are derived analytically as equations \eqref{eq:degenerate1} and \eqref{eq:degenerate2} in section \ref{sec:DegenerateRegimes}.}
    \label{fig:PhaseSpace}
\end{figure*}

After $T_u=200$ time units, we observe a notable pattern when plotting the spatial distribution of the richest or poorest 10\% of the agents (\fref{fig:Clustering_Rich_Poor_Random}). Both deciles form small clusters of neighbourhoods where they have mostly neighbours from within their own group. Using the clustering coefficient defined in section \ref{sc:MethodsandModels}, we find that in the given parameter setting, both quantiles converge to the same level of clustering and far exceed the expectation for random matrices, indicating that a significant level of clustering emerges from the system's dynamics (\fref{fig:Clustering} left panel).\\
We further find that (after normalising the autocorrelation matrix with its maximum) there are no noteworthy differences between the row-wise and column-wise autocorrelation (corresponding to the x and y directions). Both directions for both the top and bottom decile show a notably higher autocorrelation for the first nontrivial lag than the random matrix, thereby confirming the analysis of the clustering coefficients that rich and poor neighbourhoods emerge in the system (\fref{fig:Clustering} right panel).

%\section{Finding Distinct Regimes in a Parameter Scan}
\section{Parameter scan}
\label{sc:Parameterscan}
To check if the clustering observed in the previous section is a general feature of the model or an artefact of the specific parameter configuration, we perform a parameter scan across the space of $(r,c)$. For constant $N=64$ and $T_u=200$, we vary $c$ across $0, 0.05, \dots, 0.95$ and $r$ across $0, 0.05, \dots, 2$ and calculate the clustering coefficients for each pairing. We show the results in the $(r,c)$ phase space in \fref{fig:PhaseSpace}. By repeating the analysis with different values for $T_u$, we show that the qualitative behaviour is robust, though with the clustering in both regimes slowly increasing with larger $T_u$. \\  
For both the top and bottom decile, we find a gradual increase in clustering with $c$ and $r$ (\fref{fig:PhaseSpace}). However, for the bottom decile, we find a much more pronounced behaviour with a well-defined region of high clustering in the centre of the phase space. We explore this further in section~\ref{sec:DegenerateRegimes}, where we show that its borders (black lines on right panel of \fref{fig:PhaseSpace}) can be derived analytically and reflect the onset of degenerate values for the fees $F_{\min/\max}$.\\

\begin{figure}
    \centering
    \includegraphics[width = \linewidth ]{ 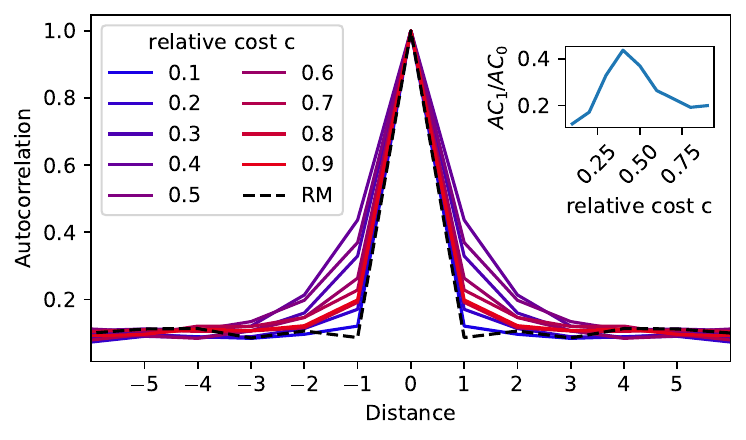}
    \caption{For constant $r = 0.5$, $c$ is varied and the autocorrelation of the poorest decile is plotted. The displayed ACFs are the mean values across the slices in the x and y directions shown in the right part of \fref{fig:Clustering}. The inset depicts the strength of the first nontrivial autocorrelation value $AC_1$ normalised by the maximum autocorrelation $AC_0$ which reaches a well-defined maximum at $c=0.4$.}
    \label{fig:ACF_scan}
\end{figure}

To visualise how sudden the clustering emerges, we scan through a slice of the phase space at $r=0.5$ by varying $c$ and plot the autocorrelation function for the top and bottom decile. Normalising the first nontrivial value of the ACF by its maximum $AC_1/AC_0$ reveals a well-defined maximum for the clustering of the bottom decile as shown in \fref{fig:ACF_scan}. On the contrary, the top decile (cf. \fref{fig:ACF_scan_rich} in the appendix) only has a linear increase of $AC_1/AC_0$ with $c$ without any indications of a discontinuity. An exploration of the highly volatile regime in the top right corner of the phase space is also given in the appendix.

\subsection{Degenerate Regimes}
\label{sec:DegenerateRegimes}
In \cite{peters2017insurance,peters_2023_editorial,ole_peters_2023_7994190}, the risk agents experience is entirely negative ($c>0$, $r = 0)$). Here, we explore a more general setup, where both $c \geq 0$ and $r \geq 0$ are varied. Hence, we also simulate regimes where the ``risk'' is favourable to agent $A$ ($r \gg c \gtrsim 0$). Hence, the question arises; how does this affect the fees?\\
The first degenerate regime is defined as $F_{\max} < 0$, i.e. the risk is so favourable for agent $A$ that they demand to receive money in exchange for giving away the risk. The border of this regime can be derived from \eref{eq:Fmax_Fmin} and is given by 
\begin{equation} 
\label{eq:degenerate1}
    c = 1-\frac{1}{1+r}.
\end{equation}
Setting the time average growth rate $g_A^{\text{without}} \overset{!}{=} 0$, i.e. having no time average growth or decline for the individual, leads to the same condition for $c$. This degeneracy line is shown as a solid black curve in the right panel of \fref{fig:PhaseSpace}. Everything below this curve results in $F_{\max}<0$.\\

Similarly, from agent $B$'s perspective, the question becomes, when is $F_{\min} < 0$, i.e. when is agent $B$ willing to pay the agent $A$ to take over the risk? Because both agents' wealth levels $w_A$ and $w_B$ are necessary to calculate $F_{\min}$ in \eqref{eq:Fmax_Fmin} and their relative difference $w_A/w_B$ can vary wildly, there exists no analytical solution to this. However, in the case of $w_B\gg w_A$ we can rewrite \eqref{eq:Fmax_Fmin} as
\begin{align}
\label{eq:degenerate2}
   0\overset{!}{=} F_{\min} &= -w_B  + \frac{1}{2} \sqrt{4w_B^2 + (c+r)^2w_A^2 } + \frac{(c-r)w_A}{2}  \\ \nonumber
    &= -w_B  + \frac{1}{2} w_B\sqrt{4 + \epsilon^2 } + \frac{(c-r)w_A}{2} \\ \nonumber
    &= -w_B + \frac{1}{2} w_B \left(2  + \mathcal{O}\left( \epsilon^2 \right)\right) + \frac{(c-r)w_A}{2} \\ \nonumber 
    &\approx \frac{(c-r)w_A}{2}\\ \nonumber 
    \Rightarrow c=r,
\end{align}
by setting $\epsilon = \frac{(c+r)w_A}{w_B}\ll 1 $ and using first-order Taylor expansion.\footnote{Note that as discussed in \cite{peters2017insurance}, when $w_A/w_B \rightarrow 0$, the time average of the risk approaches the expected value from agent $B$'s perspective. This expression is, therefore, identical to using the expected value operator.} The condition $c=r$ is shown as a dashed line in the right panel of \fref{fig:PhaseSpace}. \\
%While negative fees might seem counterintuitive at first, the model still makes sense under those conditions as explained in the discussion section \ref{subsec:Generalistion_Degenerate}.\\ % and the same relationship can also be derived by setting the ensemble average evaluation $\frac{1}{2}cw_A + \frac{1}{2}rw_A - w_A$ of the gambles to 0. \\
As seen in \fref{fig:PhaseSpace} (left panel), there is no immediate connection between the degeneracy regimes and the clustering in the top decile. However, for the poorest decile (\fref{fig:PhaseSpace}, right panel), the degeneracy lines approximate the contours of the regime of high clustering. Clustering into poor neighbourhoods, thus, happens if $c \geq 1 - \frac{1}{1+r}$ (\eqref{eq:degenerate1}) %and $F_{\max}>0$ 
and if $r > c$ (\eqref{eq:degenerate2} under the condition that $w_B\gg w_A$).

\section{Discussion}
\label{sc:Conclusion}

\begin{figure*}[t!]
        \centering
        \includegraphics[width = 0.7\textwidth]{ 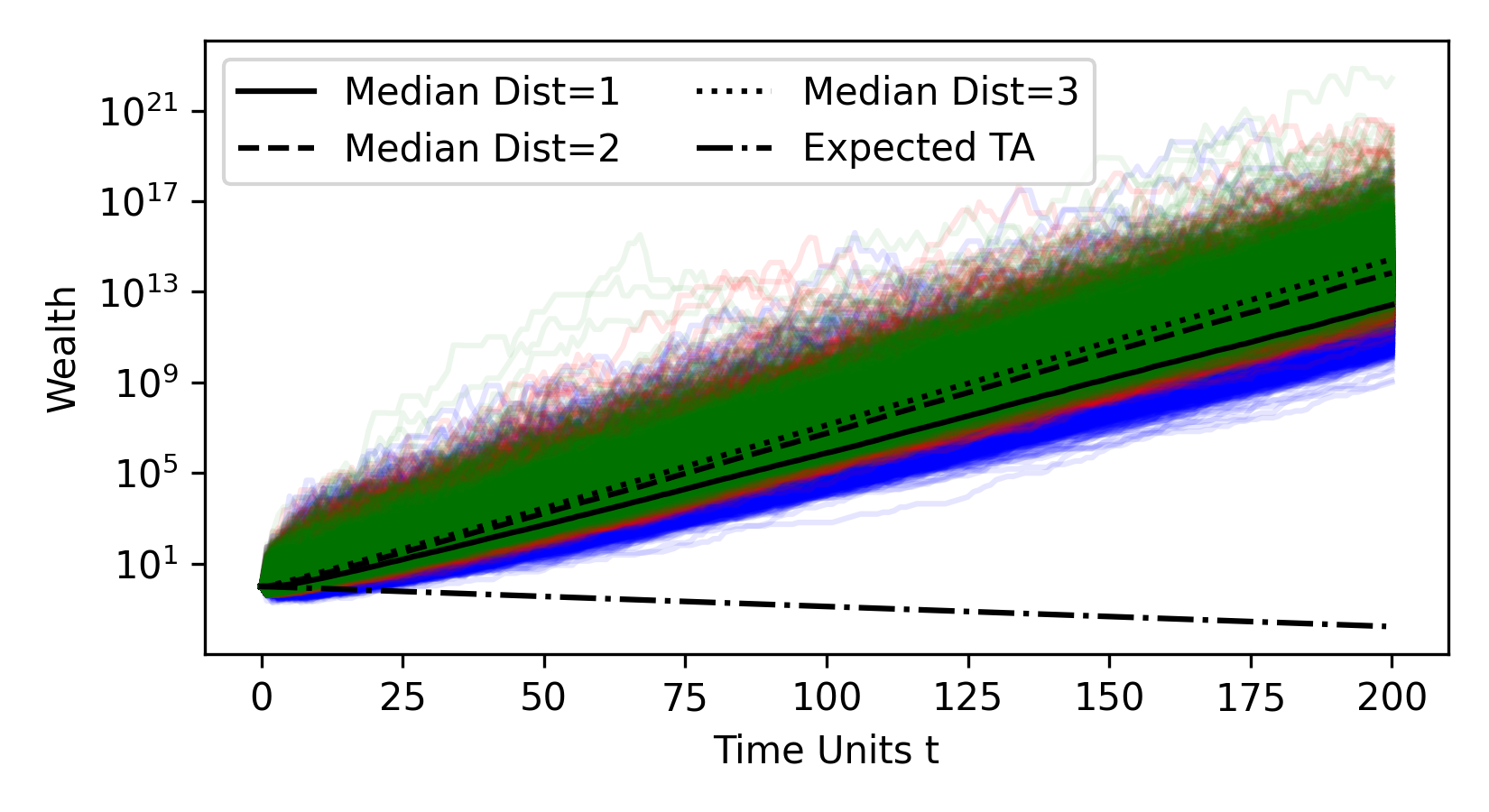}
        \caption{Wealth trajectories of the agents and their median wealth for $T=200$, $r=1.4$, $c=0.6$, $N=64$ and an interaction radius for nearest-neighbours (blue), next-nearest-neighbours (red) and next-next-nearest-neighbours (green). The median trajectories are almost the same and far outperform the expectation for the time average of any individual agent without insurance (Expected TA).}
       \label{fig:trajectories}
  \end{figure*}

\subsection{Summary}
The model presented in this paper is an extension of the agent-based model presented in \cite{peters_2023_editorial,ole_peters_2023_7994190}. We find that despite restricting the agents' ability to make contracts to their nearest neighbours on a spatially restricted grid, the overall conclusion from \cite{peters_2023_editorial}, i.e. that insurance is advantageous in the long run for all agents, still holds. 
%However, we find that the general wealth ranking is preserved for much longer periods of time, indicating that including spatial constraints can be a way to better understand the social immobility found in real societies. 
However, we find that the general wealth ranking is preserved for shorter periods of time, indicating that not having spatial constraints leads to what can be thought of as monopoly-like structures, consistent with tendencies in society with increasingly global access through the internet. We, though, highlight that the memory in the system with spatial constraints is still high, which is consistent with the well-documented poverty trap observed in real societies \cite{PovertyTrap_Ghatak2015,Persistence_inequality}. We further find that this restriction leads to a rich phenomenology of spatial cluster formations, again similar to what we see in the real world. While the clustering of the wealthiest agents increases continuously for larger parameters $r$ and $c$, we find that the poorest agents only exhibit high clustering for particular combinations of $(r,c)$ and that this regime can approximated by the degeneracy lines. %We approximate this regime by degeneracy lines, which show that the model changes from insuring a mostly harmful risk to selling a mostly positive but volatile reward.

%relevance
\subsection{Model Relevance}
While including spatial constraints to the insurance model presented in \cite{peters_2023_editorial} is an attempt to explore a more realistic system, we recognise that the system is still a very simplified representation of the world. A question we have been faced with is why we do not include specific insurance companies rather than having the insurer simply be another agent in the system. This admittedly is closer to the real world. However, this is an unnecessary restriction to the model: Even if a large insurance company exist, one still has to rely either on the non-satisfactory solutions discussed earlier or on the time solution provided here to solve the insurance paradox, and as the argument often is that such an insurance company can offer contracts as expected value (either through having such high wealth that the time average of the risk approaches the expected value or through having access to the actual ensemble), cooperation through insurance would only be stronger than what we present. We, therefore, see our model as more general.

%\subsection{Non-Ergodicity and the Advantage of Cooperation}
\subsection{Relation to Classical Cooperation}
This form of insurance can be considered a restricted form of cooperation. Although the setup on the lattice network restricts the agents in our model, the cooperation is still beneficial for them, which is in line with previous findings.
Studies on evolutionary game theory have shown that cooperative strategies can become dominant over the defectors' strategy purely due to the effect of fluctuations on the reward~\cite{PRL_Nagler_Stollmeier,StollmeierDiss}. This has also been studied within the time average framework, where cooperation is similarly found to be stable, i.e. defecting is not advantageous~\cite{PRE.Fant.PhysRevE.108.L012401}, which has been backed by empirical studies~\cite{Aktipis2016,Aktipis2011,dahl1976having}. An important note, however, on this pure form of cooperation is that trust does play a role. Despite the fact that full cooperation is mathematically stable and defecting can be shown not to be advantageous, this is only in the long run; if you give a large part of your wealth to someone in an attempt to cooperate and they decide to defect, it is, of course, not advantageous for you. This is not a consideration in our model, as both agents improve their growth rate in every time step when they sign a contract and therefore need not worry about what the other agent does in the future: taking the deal is good regardless. Whether there is a path from non-cooperation through insurance-like cooperation to full cooperation is the subject of ongoing research. Expanding the neighbourhood of potential insurance partners in an exploratory analysis indicates only minor effects on the median wealth \fref{fig:trajectories}.

\subsection{Spatial Clustering}
The main feature of this model compared to \cite{peters_2023_editorial} is the study of spatial cluster formations. Whenever the volatility is high enough, the top decile of agents tends to cluster into neighbourhoods while the bottom decile displays a similar behaviour in a specific parameter regime. The emergence of rich and poor neighbourhoods might superficially resemble kin selection~\cite{Nowak2006}. However, such a mechanism is not present in our model's algorithm. Instead, the clustering emerges endogenously: Rich agents do not choose to connect with other rich agents but rather become and stay rich because there are other rich agents in their neighbourhoods. Indeed, the original model already shows that a rich agent eventually collapses if no other agent is rich enough to insure their risk \cite{peters_2023_editorial,ole_peters_2023_7994190,Conversation}. Hence, with our spatial constraint, only agents with other rich neighbours have a chance to stay rich in the long run. This phenomenon might be interpreted as a special case of network reciprocity described by Nowak~\cite{Nowak2006}. 

\subsection{Persistent Inequality}
As shown in \fref{fig:Spearman}, the Spearman correlation with the spatial constraints is lower than the Spearman correlation found using the well-mixed model presented in \cite{peters_2023_editorial}. This is an interesting finding when we compare it to, for example, the two theories for the emergence of poverty traps discussed in \cite{PovertyTrap_Ghatak2015}: Scarcity-driven and friction-driven. The scarcity-driven poverty traps describe a different behaviour of agents under the pressure of extreme scarcity, whereas the friction-driven poverty traps describe situations in which market inefficiencies or different initial conditions lead agents with identical decision criteria to different wealth trajectories. We note that the inequality/memory in the system can be thought of as scarcity-driven: As some agents have much less wealth, the specific risk is much greater to them than a richer agent, and they are thus willing to pay a greater fee. However, this is identical whether including spatial constraints or not. Interestingly, we find the opposite to friction-driven inequality: Including spatial constraints clearly a market inefficiency, but we find that the memory of the system is greater without it. We suspect that this can be explained by monopoly-like structures which emerge in the well-mixed system but are not possible with the spatial constraint. We, though, highlight that in both our and the well-mixed system, there are no inefficiencies in forming contracts (such as administration costs), nor any agents acting in bad faith (such as agents demanding higher prices than based on their own assessment), which might change these results.

\subsection{Generalisation of the Original Model Gives Rise to Negative Fees}
\label{subsec:Generalistion_Degenerate}
Another extension of \cite{peters_2023_editorial} is to include a positive outcome of the gamble with reward $r > 0$ in addition to the harmful cost $c$. We show that this can lead to regimes where it might be mutually beneficial to have $F \leq 0$, i.e. the ``insurer'' pays an amount to take over the risk. The phase space spanned by the two parameters $(r,c)$ in \fref{fig:PhaseSpace} shows that the bottom quantile's clustering corresponds to the two degeneracy lines (\eref{eq:degenerate1} and \eref{eq:degenerate2}) derived in section \ref{sec:DegenerateRegimes}. Eq. \ref{eq:degenerate1} marks the regime at which $F_{\max}<0$ and the time average growth rate of the gamble becomes neutral, while \eref{eq:degenerate2} indicates the regime where the ensemble average becomes neutral and, under the condition $w_A\ll w_B$, $F_{\min}$ can become negative. It might be tempting to dismiss the regimes with negative fees as beyond the scope of an insurance model. However, this analysis shows that there are no qualitative differences or irregularities in the model behaviour in the different regimes:\\

First, the condition $F_{\min} < F_{\max}$ is still a meaningful model of reality. Consider $F_{\min} < 0 < F_{\max}$ as discussed above. The negative $F_{\min} < 0$ of agent $B$ means that they want to acquire the risk, while the positive $F_{\max} > 0$ of agent $A$ means that they want to get rid of it. In this scenario, a transfer at e.g. $F = 0$ (fulfilling $F_{\min} < F = 0 < F_{\max}$) is beneficial for both of them. Alternatively, if both $F_{\min} < F_{\max} < 0$, then one can more easily understand the setup by reversing the signs and considering the price $P_B = -F_{\min}$ that agent $B$ is willing to pay and $P_A = -F_{\max}$ that agent $A$ demands. Now, $0 < P_A < P_B$ and agent $B$ is willing to match agent $A$'s demanded price. The interested reader can play around with the parameters for $F_{\min} < 0 < F_{\max}$ and $F_{\min} < F_{\max} < 0$ to verify that the condition $F_{\min}<F_{\max}$ still yields sensible results. \\

Second, the regime of high clustering is characterised by $F_{\max}>0$ and the possibility that $F_{\min}<0$ if $w_A \ll w_B$, i.e. if $B$ is much wealthier than $A$, it will consider the gamble (whose reward and cost are relative to $A$'s wealth) so advantageous that $B$ offers $A$ money ($F_{\min} < 0$) to reap the potential rewards. At the same time, the risk $c$ is so large relative to $A$'s wealth that $A$ is willing to pay money in order to get rid of the risk. This is just a more extreme version of the risk assessment $0< F_{\min} < F_{\max}$ in the non-degenerate regime and illustrates that purely by considering time averages, the agents can come to mutually beneficial deals without assuming the existence of any subjective utility functions or subjective risk assessment~\cite{peters2017insurance}. Both agents operate under the same rules, but the time average considerations lead them to different risk assessments based on their individual wealth $w_{A/B}$. \\

Third, if $F_{\min} < F_{\max} < 0$, one can rethink this setup as modelling employment rather than insurance contracts. While insurances are used to replace $A$'s overall detrimental risk with a fixed payment from $A$ to $B$, one can think about employment in similar terms. Now, $A$ can work as a self-sustained freelancer and, while being mostly profitable, has varying success. Alternatively, $A$ can be hired by another agent $B$ and give $B$ the (varying) rewards of its work while, in return, getting a fixed income (a negative fee $F < 0$ corresponds to $A$ getting $-F$ from $B$). This regime is considered in more detail in \cite{ChvykovPresentation} but arises naturally from our model's parameter scan as another regime of cooperation in the face of uncertainty.

%Further work
\subsection{Further Work}
We consider this a first attempt to generalise the model presented in \cite{peters_2023_editorial}. Firstly, we find qualitatively similar results on the benefits in the long run of evaluating the risk using time average growth rates, which leaves the question of whether any non-trivial network structures lead to qualitatively different results open. Secondly, our spatial constraint reveals a decreased memory in wealth ranking. The question now is: How much does the temporal autocorrelation change if the neighbourhood of the agents is expanded or if the network structure is changed? And are there any configurations that lead to qualitatively different results? 
%Intuitively, if the agents get access to more and more potential insurance partners, we expect that market inefficiency should be reduced as they will have a larger selection of offers to choose their best deal and inequality should become less persistent, even though an exploratory analysis indicates only minor effects on the median wealth \fref{fig:trajectories}. 
Lastly, we recognise that consumption is an important feature in real systems~\cite{Lucas1978}. This leads to the possibility of bankruptcy, which in this model is equal to being erased from the system. We are aware of no solutions for dealing with such a problem on a non-ad-hoc basis, but we would be interested in discussing possible solutions. In general, we hope this paper will inspire researchers to explore the insurance problem further through the lens of time average considerations. 

\begin{acknowledgements}
The authors thank the organisers and participants of the Ergodicity Economics 2024 conference for their valuable feedback and discussions. Tobias Wand is supported by the Studienstiftung des deutschen Volkes (German Academic Scholarship Foundation). Benjamin Skjold is supported by a Novo Nordisk Exploratory Interdisciplinary Synergy grant (ref NNF20OC0064869).
\end{acknowledgements}

\bibliography{apssamp}% Produces the bibliography via BibTeX.

\begin{thebibliography}{37}%
\makeatletter
\providecommand \@ifxundefined [1]{%
 \@ifx{#1\undefined}
}%
\providecommand \@ifnum [1]{%
 \ifnum #1\expandafter \@firstoftwo
 \else \expandafter \@secondoftwo
 \fi
}%
\providecommand \@ifx [1]{%
 \ifx #1\expandafter \@firstoftwo
 \else \expandafter \@secondoftwo
 \fi
}%
\providecommand \natexlab [1]{#1}%
\providecommand \enquote  [1]{``#1''}%
\providecommand \bibnamefont  [1]{#1}%
\providecommand \bibfnamefont [1]{#1}%
\providecommand \citenamefont [1]{#1}%
\providecommand \href@noop [0]{\@secondoftwo}%
\providecommand \href [0]{\begingroup \@sanitize@url \@href}%
\providecommand \@href[1]{\@@startlink{#1}\@@href}%
\providecommand \@@href[1]{\endgroup#1\@@endlink}%
\providecommand \@sanitize@url [0]{\catcode `\\12\catcode `\$12\catcode `\&12\catcode `\#12\catcode `\^12\catcode `\_12\catcode `\%12\relax}%
\providecommand \@@startlink[1]{}%
\providecommand \@@endlink[0]{}%
\providecommand \url  [0]{\begingroup\@sanitize@url \@url }%
\providecommand \@url [1]{\endgroup\@href {#1}{\urlprefix }}%
\providecommand \urlprefix  [0]{URL }%
\providecommand \Eprint [0]{\href }%
\providecommand \doibase [0]{https://doi.org/}%
\providecommand \selectlanguage [0]{\@gobble}%
\providecommand \bibinfo  [0]{\@secondoftwo}%
\providecommand \bibfield  [0]{\@secondoftwo}%
\providecommand \translation [1]{[#1]}%
\providecommand \BibitemOpen [0]{}%
\providecommand \bibitemStop [0]{}%
\providecommand \bibitemNoStop [0]{.\EOS\space}%
\providecommand \EOS [0]{\spacefactor3000\relax}%
\providecommand \BibitemShut  [1]{\csname bibitem#1\endcsname}%
\let\auto@bib@innerbib\@empty
%</preamble>
\bibitem [{\citenamefont {Zhang}(1991)}]{Zhang1991}%
  \BibitemOpen
  \bibfield  {author} {\bibinfo {author} {\bibfnamefont {W.-B.}\ \bibnamefont {Zhang}},\ }\href {https://doi.org/10.1007/978-3-642-75909-3} {\emph {\bibinfo {title} {Synergetic Economics}}}\ (\bibinfo  {publisher} {Springer Berlin Heidelberg},\ \bibinfo {year} {1991})\BibitemShut {NoStop}%
\bibitem [{\citenamefont {Allen}(2007)}]{selforganisation_economic}%
  \BibitemOpen
  \bibfield  {author} {\bibinfo {author} {\bibfnamefont {P.}~\bibnamefont {Allen}},\ }\bibinfo {title} {Self-organization in economic systems}\ (\bibinfo  {publisher} {Edward Elgar Publishing},\ \bibinfo {year} {2007})\ pp.\ \bibinfo {pages} {1111--1148}\BibitemShut {NoStop}%
\bibitem [{\citenamefont {Strogatz}(2000)}]{strogatz_book}%
  \BibitemOpen
  \bibfield  {author} {\bibinfo {author} {\bibfnamefont {S.}~\bibnamefont {Strogatz}},\ }\href@noop {} {\emph {\bibinfo {title} {Nonlinear Dynamics and Chaos: With Applications to Physics, Biology, Chemistry and Engineering}}},\ Studies in nonlinearity\ (\bibinfo  {publisher} {Westview},\ \bibinfo {year} {2000})\BibitemShut {NoStop}%
\bibitem [{\citenamefont {Foster}(1993)}]{selforganisation_economics_irreversibility_entropy}%
  \BibitemOpen
  \bibfield  {author} {\bibinfo {author} {\bibfnamefont {J.}~\bibnamefont {Foster}},\ }\bibfield  {title} {\bibinfo {title} {{Economics and the Self-Organisation Approach: Alfred Marshall Revisited?}},\ }\href {https://doi.org/10.2307/2234714} {\bibfield  {journal} {\bibinfo  {journal} {The Economic Journal}\ }\textbf {\bibinfo {volume} {103}},\ \bibinfo {pages} {975} (\bibinfo {year} {1993})},\ \Eprint {https://arxiv.org/abs/https://academic.oup.com/ej/article-pdf/103/419/975/27042593/ej0975.pdf} {https://academic.oup.com/ej/article-pdf/103/419/975/27042593/ej0975.pdf} \BibitemShut {NoStop}%
\bibitem [{\citenamefont {Peters}\ and\ \citenamefont {Adamou}(2017)}]{peters2017insurance}%
  \BibitemOpen
  \bibfield  {author} {\bibinfo {author} {\bibfnamefont {O.}~\bibnamefont {Peters}}\ and\ \bibinfo {author} {\bibfnamefont {A.}~\bibnamefont {Adamou}},\ }\href@noop {} {\bibinfo {title} {Insurance makes wealth grow faster}} (\bibinfo {year} {2017}),\ \Eprint {https://arxiv.org/abs/1507.04655} {arXiv:1507.04655 [q-fin.RM]} \BibitemShut {NoStop}%
\bibitem [{\citenamefont {Peters}(2023)}]{peters_2023_editorial}%
  \BibitemOpen
  \bibfield  {author} {\bibinfo {author} {\bibfnamefont {O.}~\bibnamefont {Peters}},\ }\bibfield  {title} {\bibinfo {title} {Insurance as an ergodicity problem},\ }\href {https://doi.org/10.1017/S1748499523000131} {\bibfield  {journal} {\bibinfo  {journal} {Annals of Actuarial Science}\ }\textbf {\bibinfo {volume} {17}},\ \bibinfo {pages} {215–218} (\bibinfo {year} {2023})}\BibitemShut {NoStop}%
\bibitem [{\citenamefont {Peters}\ and\ \citenamefont {Skjold}(2023)}]{ole_peters_2023_7994190}%
  \BibitemOpen
  \bibfield  {author} {\bibinfo {author} {\bibfnamefont {O.}~\bibnamefont {Peters}}\ and\ \bibinfo {author} {\bibfnamefont {B.}~\bibnamefont {Skjold}},\ }\href {https://doi.org/10.5281/zenodo.7994190} {\bibinfo {title} {{LMLhub/AAS\_editorial\_figures\_July2023: v1.0.0 initial release}}} (\bibinfo {year} {2023}),\ \bibinfo {note} {code published on zenodo}\BibitemShut {NoStop}%
\bibitem [{\citenamefont {Ising}(1925)}]{Ising1925}%
  \BibitemOpen
  \bibfield  {author} {\bibinfo {author} {\bibfnamefont {E.}~\bibnamefont {Ising}},\ }\bibfield  {title} {\bibinfo {title} {Beitrag zur theorie des ferromagnetismus},\ }\href {https://doi.org/10.1007/bf02980577} {\bibfield  {journal} {\bibinfo  {journal} {Zeitschrift f\"{u}r Physik}\ }\textbf {\bibinfo {volume} {31}},\ \bibinfo {pages} {253–258} (\bibinfo {year} {1925})}\BibitemShut {NoStop}%
\bibitem [{\citenamefont {Onsager}(1944)}]{Onsager_Ising_PhysRev.65.117}%
  \BibitemOpen
  \bibfield  {author} {\bibinfo {author} {\bibfnamefont {L.}~\bibnamefont {Onsager}},\ }\bibfield  {title} {\bibinfo {title} {Crystal statistics. i. a two-dimensional model with an order-disorder transition},\ }\href {https://doi.org/10.1103/PhysRev.65.117} {\bibfield  {journal} {\bibinfo  {journal} {Phys. Rev.}\ }\textbf {\bibinfo {volume} {65}},\ \bibinfo {pages} {117} (\bibinfo {year} {1944})}\BibitemShut {NoStop}%
\bibitem [{\citenamefont {Nowak}\ and\ \citenamefont {May}(1992)}]{Nowak1992}%
  \BibitemOpen
  \bibfield  {author} {\bibinfo {author} {\bibfnamefont {M.~A.}\ \bibnamefont {Nowak}}\ and\ \bibinfo {author} {\bibfnamefont {R.~M.}\ \bibnamefont {May}},\ }\bibfield  {title} {\bibinfo {title} {Evolutionary games and spatial chaos},\ }\href {https://doi.org/10.1038/359826a0} {\bibfield  {journal} {\bibinfo  {journal} {Nature}\ }\textbf {\bibinfo {volume} {359}},\ \bibinfo {pages} {826–829} (\bibinfo {year} {1992})}\BibitemShut {NoStop}%
\bibitem [{\citenamefont {Sznajd-Weron}\ and\ \citenamefont {Sznajd}(2000)}]{Ising-like_SZNAJD-WERON}%
  \BibitemOpen
  \bibfield  {author} {\bibinfo {author} {\bibfnamefont {K.}~\bibnamefont {Sznajd-Weron}}\ and\ \bibinfo {author} {\bibfnamefont {J.}~\bibnamefont {Sznajd}},\ }\bibfield  {title} {\bibinfo {title} {Opinion evolution in closed community},\ }\href {https://doi.org/10.1142/S0129183100000936} {\bibfield  {journal} {\bibinfo  {journal} {International Journal of Modern Physics C}\ }\textbf {\bibinfo {volume} {11}},\ \bibinfo {pages} {1157} (\bibinfo {year} {2000})}\BibitemShut {NoStop}%
\bibitem [{\citenamefont {Stauffer}(2008)}]{Stauffer2008}%
  \BibitemOpen
  \bibfield  {author} {\bibinfo {author} {\bibfnamefont {D.}~\bibnamefont {Stauffer}},\ }\bibfield  {title} {\bibinfo {title} {Social applications of two-dimensional ising models},\ }\href {https://doi.org/10.1119/1.2779882} {\bibfield  {journal} {\bibinfo  {journal} {American Journal of Physics}\ }\textbf {\bibinfo {volume} {76}},\ \bibinfo {pages} {470–473} (\bibinfo {year} {2008})}\BibitemShut {NoStop}%
\bibitem [{\citenamefont {Sornette}(2014)}]{Sornette2014}%
  \BibitemOpen
  \bibfield  {author} {\bibinfo {author} {\bibfnamefont {D.}~\bibnamefont {Sornette}},\ }\bibfield  {title} {\bibinfo {title} {Physics and financial economics (1776–2014): puzzles, ising and agent-based models},\ }\href {https://doi.org/10.1088/0034-4885/77/6/062001} {\bibfield  {journal} {\bibinfo  {journal} {Reports on Progress in Physics}\ }\textbf {\bibinfo {volume} {77}},\ \bibinfo {pages} {062001} (\bibinfo {year} {2014})}\BibitemShut {NoStop}%
\bibitem [{\citenamefont {Schelling}(1971)}]{SegregationModel}%
  \BibitemOpen
  \bibfield  {author} {\bibinfo {author} {\bibfnamefont {T.~C.}\ \bibnamefont {Schelling}},\ }\bibfield  {title} {\bibinfo {title} {Dynamic models of segregation},\ }\href {https://doi.org/10.1080/0022250X.1971.9989794} {\bibfield  {journal} {\bibinfo  {journal} {The Journal of Mathematical Sociology}\ }\textbf {\bibinfo {volume} {1}},\ \bibinfo {pages} {143} (\bibinfo {year} {1971})}\BibitemShut {NoStop}%
\bibitem [{\citenamefont {Peters}\ and\ \citenamefont {Adamou}(2018)}]{Script}%
  \BibitemOpen
  \bibfield  {author} {\bibinfo {author} {\bibfnamefont {O.}~\bibnamefont {Peters}}\ and\ \bibinfo {author} {\bibfnamefont {A.}~\bibnamefont {Adamou}},\ }\href@noop {} {\bibinfo {title} {Ergodicity economics lecture notes}} (\bibinfo {year} {2018})\BibitemShut {NoStop}%
\bibitem [{\citenamefont {Peters}\ and\ \citenamefont {Gell-Mann}(2016)}]{Peters_Gell-Mann}%
  \BibitemOpen
  \bibfield  {author} {\bibinfo {author} {\bibfnamefont {O.}~\bibnamefont {Peters}}\ and\ \bibinfo {author} {\bibfnamefont {M.}~\bibnamefont {Gell-Mann}},\ }\bibfield  {title} {\bibinfo {title} {{Evaluating gambles using dynamics}},\ }\href {https://doi.org/10.1063/1.4940236} {\bibfield  {journal} {\bibinfo  {journal} {Chaos: An Interdisciplinary Journal of Nonlinear Science}\ }\textbf {\bibinfo {volume} {26}},\ \bibinfo {pages} {023103} (\bibinfo {year} {2016})}\BibitemShut {NoStop}%
\bibitem [{\citenamefont {Chiappori}\ \emph {et~al.}(2006)\citenamefont {Chiappori}, \citenamefont {Jullien}, \citenamefont {Salanié},\ and\ \citenamefont {Salanié}}]{AsymmetricInformation_RiskAversion}%
  \BibitemOpen
  \bibfield  {author} {\bibinfo {author} {\bibfnamefont {P.-A.}\ \bibnamefont {Chiappori}}, \bibinfo {author} {\bibfnamefont {B.}~\bibnamefont {Jullien}}, \bibinfo {author} {\bibfnamefont {B.}~\bibnamefont {Salanié}},\ and\ \bibinfo {author} {\bibfnamefont {F.}~\bibnamefont {Salanié}},\ }\bibfield  {title} {\bibinfo {title} {Asymmetric information in insurance: General testable implications},\ }\href {http://www.jstor.org/stable/25046274} {\bibfield  {journal} {\bibinfo  {journal} {The RAND Journal of Economics}\ }\textbf {\bibinfo {volume} {37}},\ \bibinfo {pages} {783} (\bibinfo {year} {2006})}\BibitemShut {NoStop}%
\bibitem [{\citenamefont {Arrow}(1971)}]{arrow1971theory}%
  \BibitemOpen
  \bibfield  {author} {\bibinfo {author} {\bibfnamefont {K.~J.}\ \bibnamefont {Arrow}},\ }\bibfield  {title} {\bibinfo {title} {The theory of risk aversion},\ }\href@noop {} {\bibfield  {journal} {\bibinfo  {journal} {Essays in the theory of risk-bearing}\ ,\ \bibinfo {pages} {90}} (\bibinfo {year} {1971})}\BibitemShut {NoStop}%
\bibitem [{\citenamefont {Peters}\ and\ \citenamefont {Adamou}(2022)}]{Peters2022_cooperation}%
  \BibitemOpen
  \bibfield  {author} {\bibinfo {author} {\bibfnamefont {O.}~\bibnamefont {Peters}}\ and\ \bibinfo {author} {\bibfnamefont {A.}~\bibnamefont {Adamou}},\ }\bibfield  {title} {\bibinfo {title} {The ergodicity solution of the cooperation puzzle},\ }\bibfield  {journal} {\bibinfo  {journal} {Philosophical Transactions of the Royal Society A: Mathematical, Physical and Engineering Sciences}\ }\textbf {\bibinfo {volume} {380}},\ \href {https://doi.org/10.1098/rsta.2020.0425} {10.1098/rsta.2020.0425} (\bibinfo {year} {2022})\BibitemShut {NoStop}%
\bibitem [{\citenamefont {Meder}\ \emph {et~al.}(2021)\citenamefont {Meder}, \citenamefont {Rabe}, \citenamefont {Morville}, \citenamefont {Madsen}, \citenamefont {Koudahl}, \citenamefont {Dolan}, \citenamefont {Siebner},\ and\ \citenamefont {Hulme}}]{CopenhagenExperiment}%
  \BibitemOpen
  \bibfield  {author} {\bibinfo {author} {\bibfnamefont {D.}~\bibnamefont {Meder}}, \bibinfo {author} {\bibfnamefont {F.}~\bibnamefont {Rabe}}, \bibinfo {author} {\bibfnamefont {T.}~\bibnamefont {Morville}}, \bibinfo {author} {\bibfnamefont {K.~H.}\ \bibnamefont {Madsen}}, \bibinfo {author} {\bibfnamefont {M.~T.}\ \bibnamefont {Koudahl}}, \bibinfo {author} {\bibfnamefont {R.~J.}\ \bibnamefont {Dolan}}, \bibinfo {author} {\bibfnamefont {H.~R.}\ \bibnamefont {Siebner}},\ and\ \bibinfo {author} {\bibfnamefont {O.~J.}\ \bibnamefont {Hulme}},\ }\bibfield  {title} {\bibinfo {title} {Ergodicity-breaking reveals time optimal decision making in humans},\ }\href {https://doi.org/10.1371/journal.pcbi.1009217} {\bibfield  {journal} {\bibinfo  {journal} {PLOS Computational Biology}\ }\textbf {\bibinfo {volume} {17}},\ \bibinfo {pages} {1} (\bibinfo {year} {2021})}\BibitemShut {NoStop}%
\bibitem [{\citenamefont {Vanhoyweghen}\ and\ \citenamefont {Ginis}(2023)}]{Vanhoyweghen2023}%
  \BibitemOpen
  \bibfield  {author} {\bibinfo {author} {\bibfnamefont {A.}~\bibnamefont {Vanhoyweghen}}\ and\ \bibinfo {author} {\bibfnamefont {V.}~\bibnamefont {Ginis}},\ }\bibfield  {title} {\bibinfo {title} {Human decision-making in a non-ergodic additive environment},\ }\bibfield  {journal} {\bibinfo  {journal} {Proceedings of the Royal Society A: Mathematical, Physical and Engineering Sciences}\ }\textbf {\bibinfo {volume} {479}},\ \href {https://doi.org/10.1098/rspa.2023.0544} {10.1098/rspa.2023.0544} (\bibinfo {year} {2023})\BibitemShut {NoStop}%
\bibitem [{\citenamefont {Skjold}\ \emph {et~al.}(2023)\citenamefont {Skjold}, \citenamefont {Steinkamp}, \citenamefont {Connaughton}, \citenamefont {Hulme},\ and\ \citenamefont {Peters}}]{Skjold2023}%
  \BibitemOpen
  \bibfield  {author} {\bibinfo {author} {\bibfnamefont {B.}~\bibnamefont {Skjold}}, \bibinfo {author} {\bibfnamefont {S.~R.}\ \bibnamefont {Steinkamp}}, \bibinfo {author} {\bibfnamefont {C.}~\bibnamefont {Connaughton}}, \bibinfo {author} {\bibfnamefont {O.~J.}\ \bibnamefont {Hulme}},\ and\ \bibinfo {author} {\bibfnamefont {O.}~\bibnamefont {Peters}},\ }\bibfield  {title} {\bibinfo {title} {Are risk preferences optimal?},\ }\bibfield  {journal} {\bibinfo  {journal} {Preprint available at}\ }\href {https://doi.org/10.31219/osf.io/ew2sx} {10.31219/osf.io/ew2sx} (\bibinfo {year} {2023})\BibitemShut {NoStop}%
\bibitem [{\citenamefont {Skjold}\ and\ \citenamefont {Peters}(2023{\natexlab{a}})}]{InsuranceModelBlogPost}%
  \BibitemOpen
  \bibfield  {author} {\bibinfo {author} {\bibfnamefont {B.}~\bibnamefont {Skjold}}\ and\ \bibinfo {author} {\bibfnamefont {O.}~\bibnamefont {Peters}},\ }\href@noop {} {\bibinfo {title} {{Insurance as an ergodicity problem }}},\ \bibinfo {howpublished} {\url{https://ergodicityeconomics.com/2023/08/08/insurance-as-an-ergodicity-problem/}} (\bibinfo {year} {2023}{\natexlab{a}}),\ \bibinfo {note} {[Online; accessed 11-August-2023]}\BibitemShut {NoStop}%
\bibitem [{Cod()}]{Code}%
  \BibitemOpen
  \href@noop {} {}\bibinfo {note} {Upon acceptance of the article, our simulation code will be made available on zenodo.}\BibitemShut {Stop}%
\bibitem [{\citenamefont {Virtanen}\ \emph {et~al.}(2020)\citenamefont {Virtanen}, \citenamefont {Gommers}, \citenamefont {Oliphant}, \citenamefont {Haberland}, \citenamefont {Reddy}, \citenamefont {Cournapeau}, \citenamefont {Burovski}, \citenamefont {Peterson}, \citenamefont {Weckesser}, \citenamefont {Bright}, \citenamefont {{van der Walt}}, \citenamefont {Brett}, \citenamefont {Wilson}, \citenamefont {Millman}, \citenamefont {Mayorov}, \citenamefont {Nelson}, \citenamefont {Jones}, \citenamefont {Kern}, \citenamefont {Larson}, \citenamefont {Carey}, \citenamefont {Polat}, \citenamefont {Feng}, \citenamefont {Moore}, \citenamefont {{VanderPlas}}, \citenamefont {Laxalde}, \citenamefont {Perktold}, \citenamefont {Cimrman}, \citenamefont {Henriksen}, \citenamefont {Quintero}, \citenamefont {Harris}, \citenamefont {Archibald}, \citenamefont {Ribeiro}, \citenamefont {Pedregosa}, \citenamefont {{van Mulbregt}},\ and\ \citenamefont {{SciPy 1.0 Contributors}}}]{2020SciPy-NMeth}%
  \BibitemOpen
  \bibfield  {author} {\bibinfo {author} {\bibfnamefont {P.}~\bibnamefont {Virtanen}}, \bibinfo {author} {\bibfnamefont {R.}~\bibnamefont {Gommers}}, \bibinfo {author} {\bibfnamefont {T.~E.}\ \bibnamefont {Oliphant}}, \bibinfo {author} {\bibfnamefont {M.}~\bibnamefont {Haberland}}, \bibinfo {author} {\bibfnamefont {T.}~\bibnamefont {Reddy}}, \bibinfo {author} {\bibfnamefont {D.}~\bibnamefont {Cournapeau}}, \bibinfo {author} {\bibfnamefont {E.}~\bibnamefont {Burovski}}, \bibinfo {author} {\bibfnamefont {P.}~\bibnamefont {Peterson}}, \bibinfo {author} {\bibfnamefont {W.}~\bibnamefont {Weckesser}}, \bibinfo {author} {\bibfnamefont {J.}~\bibnamefont {Bright}}, \bibinfo {author} {\bibfnamefont {S.~J.}\ \bibnamefont {{van der Walt}}}, \bibinfo {author} {\bibfnamefont {M.}~\bibnamefont {Brett}}, \bibinfo {author} {\bibfnamefont {J.}~\bibnamefont {Wilson}}, \bibinfo {author} {\bibfnamefont {K.~J.}\ \bibnamefont {Millman}}, \bibinfo {author} {\bibfnamefont {N.}~\bibnamefont {Mayorov}}, \bibinfo {author} {\bibfnamefont
  {A.~R.~J.}\ \bibnamefont {Nelson}}, \bibinfo {author} {\bibfnamefont {E.}~\bibnamefont {Jones}}, \bibinfo {author} {\bibfnamefont {R.}~\bibnamefont {Kern}}, \bibinfo {author} {\bibfnamefont {E.}~\bibnamefont {Larson}}, \bibinfo {author} {\bibfnamefont {C.~J.}\ \bibnamefont {Carey}}, \bibinfo {author} {\bibfnamefont {{\.I}.}~\bibnamefont {Polat}}, \bibinfo {author} {\bibfnamefont {Y.}~\bibnamefont {Feng}}, \bibinfo {author} {\bibfnamefont {E.~W.}\ \bibnamefont {Moore}}, \bibinfo {author} {\bibfnamefont {J.}~\bibnamefont {{VanderPlas}}}, \bibinfo {author} {\bibfnamefont {D.}~\bibnamefont {Laxalde}}, \bibinfo {author} {\bibfnamefont {J.}~\bibnamefont {Perktold}}, \bibinfo {author} {\bibfnamefont {R.}~\bibnamefont {Cimrman}}, \bibinfo {author} {\bibfnamefont {I.}~\bibnamefont {Henriksen}}, \bibinfo {author} {\bibfnamefont {E.~A.}\ \bibnamefont {Quintero}}, \bibinfo {author} {\bibfnamefont {C.~R.}\ \bibnamefont {Harris}}, \bibinfo {author} {\bibfnamefont {A.~M.}\ \bibnamefont {Archibald}}, \bibinfo {author}
  {\bibfnamefont {A.~H.}\ \bibnamefont {Ribeiro}}, \bibinfo {author} {\bibfnamefont {F.}~\bibnamefont {Pedregosa}}, \bibinfo {author} {\bibfnamefont {P.}~\bibnamefont {{van Mulbregt}}},\ and\ \bibinfo {author} {\bibnamefont {{SciPy 1.0 Contributors}}},\ }\bibfield  {title} {\bibinfo {title} {{{SciPy} 1.0: Fundamental Algorithms for Scientific Computing in Python}},\ }\href {https://doi.org/10.1038/s41592-019-0686-2} {\bibfield  {journal} {\bibinfo  {journal} {Nature Methods}\ }\textbf {\bibinfo {volume} {17}},\ \bibinfo {pages} {261} (\bibinfo {year} {2020})}\BibitemShut {NoStop}%
\bibitem [{\citenamefont {Ghatak}(2015)}]{PovertyTrap_Ghatak2015}%
  \BibitemOpen
  \bibfield  {author} {\bibinfo {author} {\bibfnamefont {M.}~\bibnamefont {Ghatak}},\ }\bibfield  {title} {\bibinfo {title} {Theories of poverty traps and anti-poverty policies},\ }\href {https://doi.org/10.1093/wber/lhv021} {\bibfield  {journal} {\bibinfo  {journal} {The World Bank Economic Review}\ }\textbf {\bibinfo {volume} {29}},\ \bibinfo {pages} {S77–S105} (\bibinfo {year} {2015})}\BibitemShut {NoStop}%
\bibitem [{\citenamefont {Boix}(2010)}]{Persistence_inequality}%
  \BibitemOpen
  \bibfield  {author} {\bibinfo {author} {\bibfnamefont {C.}~\bibnamefont {Boix}},\ }\bibfield  {title} {\bibinfo {title} {Origins and persistence of economic inequality},\ }\href {https://doi.org/10.1146/annurev.polisci.12.031607.094915} {\bibfield  {journal} {\bibinfo  {journal} {Annual Review of Political Science}\ }\textbf {\bibinfo {volume} {13}},\ \bibinfo {pages} {489} (\bibinfo {year} {2010})}\BibitemShut {NoStop}%
\bibitem [{\citenamefont {Stollmeier}\ and\ \citenamefont {Nagler}(2018)}]{PRL_Nagler_Stollmeier}%
  \BibitemOpen
  \bibfield  {author} {\bibinfo {author} {\bibfnamefont {F.}~\bibnamefont {Stollmeier}}\ and\ \bibinfo {author} {\bibfnamefont {J.}~\bibnamefont {Nagler}},\ }\bibfield  {title} {\bibinfo {title} {Unfair and anomalous evolutionary dynamics from fluctuating payoffs},\ }\href {https://doi.org/10.1103/PhysRevLett.120.058101} {\bibfield  {journal} {\bibinfo  {journal} {Phys. Rev. Lett.}\ }\textbf {\bibinfo {volume} {120}},\ \bibinfo {pages} {058101} (\bibinfo {year} {2018})}\BibitemShut {NoStop}%
\bibitem [{\citenamefont {Stollmeier}(2018)}]{StollmeierDiss}%
  \BibitemOpen
  \bibfield  {author} {\bibinfo {author} {\bibfnamefont {F.}~\bibnamefont {Stollmeier}},\ }\emph {\bibinfo {title} {Evolutionary Dynamics in Changing Environments}},\ \href {https://doi.org/10.53846/goediss-6920} {Ph.D. thesis},\ \bibinfo  {school} {Georg-August-Universität Göttingen} (\bibinfo {year} {2018})\BibitemShut {NoStop}%
\bibitem [{\citenamefont {Fant}\ \emph {et~al.}(2023)\citenamefont {Fant}, \citenamefont {Mazzarisi}, \citenamefont {Panizon},\ and\ \citenamefont {Grilli}}]{PRE.Fant.PhysRevE.108.L012401}%
  \BibitemOpen
  \bibfield  {author} {\bibinfo {author} {\bibfnamefont {L.}~\bibnamefont {Fant}}, \bibinfo {author} {\bibfnamefont {O.}~\bibnamefont {Mazzarisi}}, \bibinfo {author} {\bibfnamefont {E.}~\bibnamefont {Panizon}},\ and\ \bibinfo {author} {\bibfnamefont {J.}~\bibnamefont {Grilli}},\ }\bibfield  {title} {\bibinfo {title} {Stable cooperation emerges in stochastic multiplicative growth},\ }\href {https://doi.org/10.1103/PhysRevE.108.L012401} {\bibfield  {journal} {\bibinfo  {journal} {Phys. Rev. E}\ }\textbf {\bibinfo {volume} {108}},\ \bibinfo {pages} {L012401} (\bibinfo {year} {2023})}\BibitemShut {NoStop}%
\bibitem [{\citenamefont {Aktipis}\ \emph {et~al.}(2016)\citenamefont {Aktipis}, \citenamefont {de~Aguiar}, \citenamefont {Flaherty}, \citenamefont {Iyer}, \citenamefont {Sonkoi},\ and\ \citenamefont {Cronk}}]{Aktipis2016}%
  \BibitemOpen
  \bibfield  {author} {\bibinfo {author} {\bibfnamefont {A.}~\bibnamefont {Aktipis}}, \bibinfo {author} {\bibfnamefont {R.}~\bibnamefont {de~Aguiar}}, \bibinfo {author} {\bibfnamefont {A.}~\bibnamefont {Flaherty}}, \bibinfo {author} {\bibfnamefont {P.}~\bibnamefont {Iyer}}, \bibinfo {author} {\bibfnamefont {D.}~\bibnamefont {Sonkoi}},\ and\ \bibinfo {author} {\bibfnamefont {L.}~\bibnamefont {Cronk}},\ }\bibfield  {title} {\bibinfo {title} {Cooperation in an uncertain world: For the maasai of east africa, need-based transfers outperform account-keeping in volatile environments},\ }\href {https://doi.org/10.1007/s10745-016-9823-z} {\bibfield  {journal} {\bibinfo  {journal} {Human Ecology}\ }\textbf {\bibinfo {volume} {44}},\ \bibinfo {pages} {353} (\bibinfo {year} {2016})}\BibitemShut {NoStop}%
\bibitem [{\citenamefont {Aktipis}\ \emph {et~al.}(2011)\citenamefont {Aktipis}, \citenamefont {Cronk},\ and\ \citenamefont {de~Aguiar}}]{Aktipis2011}%
  \BibitemOpen
  \bibfield  {author} {\bibinfo {author} {\bibfnamefont {C.~A.}\ \bibnamefont {Aktipis}}, \bibinfo {author} {\bibfnamefont {L.}~\bibnamefont {Cronk}},\ and\ \bibinfo {author} {\bibfnamefont {R.}~\bibnamefont {de~Aguiar}},\ }\bibfield  {title} {\bibinfo {title} {Risk-pooling and herd survival: An agent-based model of a maasai gift-giving system},\ }\href {https://doi.org/10.1007/s10745-010-9364-9} {\bibfield  {journal} {\bibinfo  {journal} {Human Ecology}\ }\textbf {\bibinfo {volume} {39}},\ \bibinfo {pages} {131} (\bibinfo {year} {2011})}\BibitemShut {NoStop}%
\bibitem [{\citenamefont {Dahl}\ and\ \citenamefont {Hjort}(1976)}]{dahl1976having}%
  \BibitemOpen
  \bibfield  {author} {\bibinfo {author} {\bibfnamefont {G.}~\bibnamefont {Dahl}}\ and\ \bibinfo {author} {\bibfnamefont {A.}~\bibnamefont {Hjort}},\ }\href@noop {} {\emph {\bibinfo {title} {Having herds: Pastoral herd growth and household economy}}}\ (\bibinfo  {publisher} {Department of Social Anthropology, University of Stockholm},\ \bibinfo {year} {1976})\BibitemShut {NoStop}%
\bibitem [{\citenamefont {Nowak}(2006)}]{Nowak2006}%
  \BibitemOpen
  \bibfield  {author} {\bibinfo {author} {\bibfnamefont {M.~A.}\ \bibnamefont {Nowak}},\ }\bibfield  {title} {\bibinfo {title} {Five rules for the evolution of cooperation},\ }\href {https://doi.org/10.1126/science.1133755} {\bibfield  {journal} {\bibinfo  {journal} {Science}\ }\textbf {\bibinfo {volume} {314}},\ \bibinfo {pages} {1560–1563} (\bibinfo {year} {2006})}\BibitemShut {NoStop}%
\bibitem [{\citenamefont {Skjold}\ and\ \citenamefont {Peters}(2023{\natexlab{b}})}]{Conversation}%
  \BibitemOpen
  \bibfield  {author} {\bibinfo {author} {\bibfnamefont {B.}~\bibnamefont {Skjold}}\ and\ \bibinfo {author} {\bibfnamefont {O.}~\bibnamefont {Peters}},\ }\href@noop {} {}\bibinfo {howpublished} {Personal communication} (\bibinfo {year} {2023}{\natexlab{b}})\BibitemShut {NoStop}%
\bibitem [{\citenamefont {Chvykov}\ and\ \citenamefont {Warren}(2024)}]{ChvykovPresentation}%
  \BibitemOpen
  \bibfield  {author} {\bibinfo {author} {\bibfnamefont {P.}~\bibnamefont {Chvykov}}\ and\ \bibinfo {author} {\bibfnamefont {J.~N.}\ \bibnamefont {Warren}},\ }\bibfield  {title} {\bibinfo {title} {Is salary just a negative insurance premium?}} (\bibinfo {year} {2024}),\ \bibinfo {note} {presentation given at Ergodicity Economics Conference}\BibitemShut {NoStop}%
\bibitem [{\citenamefont {Lucas}(1978)}]{Lucas1978}%
  \BibitemOpen
  \bibfield  {author} {\bibinfo {author} {\bibfnamefont {R.~E.}\ \bibnamefont {Lucas}},\ }\bibfield  {title} {\bibinfo {title} {Asset prices in an exchange economy},\ }\href {https://doi.org/10.2307/1913837} {\bibfield  {journal} {\bibinfo  {journal} {Econometrica}\ }\textbf {\bibinfo {volume} {46}},\ \bibinfo {pages} {1429} (\bibinfo {year} {1978})}\BibitemShut {NoStop}%
\end{thebibliography}%

\subsection*{Appendix: ACF for the Richest Decile}%\huge Appendix}
\begin{figure}[h!]
    \centering
    \includegraphics[width = 0.9\linewidth ]{ 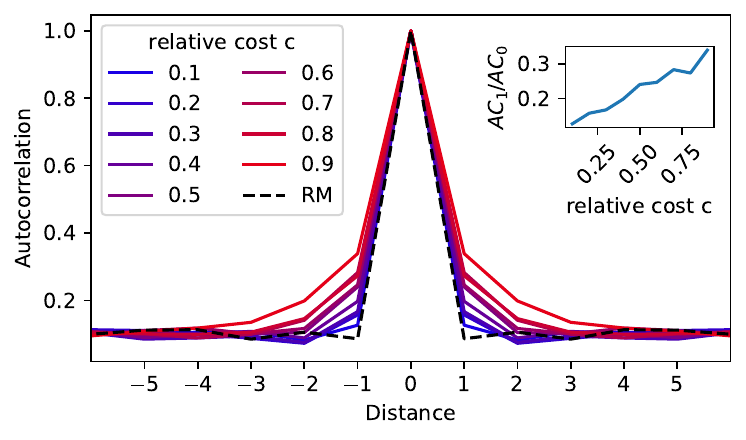}
    \caption{The same plot as \fref{fig:ACF_scan} but for the richest decile. The ACF seems to continuously increase with rising $c$.}
    \label{fig:ACF_scan_rich}
\end{figure}

\subsection*{Appendix: Highly Volatile Regime}
\begin{figure}[h]
    \centering
   \includegraphics[width = 0.99\linewidth ]{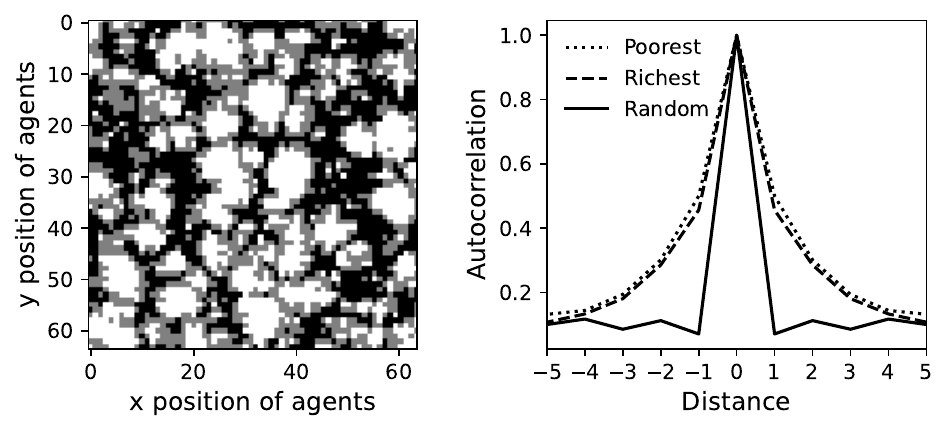}
    \caption{Results after $T=200$ time units for $N=64$ in the highly volatile regime of $c=0.95$ and $r=2$. \textbf{Left:} The colour coding now shows the agents divided into three categories as the richest (white), central (grey) and poorest thirds (black). \textbf{Right:} The ACF of the richest and poorest decile now shows much wider tails than in the previous parameter setting in \fref{fig:Clustering}.}
    \label{fig:HighlyVolatile}
\end{figure}

The phase space in \fref{fig:PhaseSpace} reveals that a particularly high clustering for both the top and bottom quantile can be found in a region of high $c$ and high $r$, i.e. where the volatility of the gamble is very large in both directions. While the low-volatile regime transitions into the deterministic case, it is not obvious how the model behaves in the high-volatile regime. To investigate this, we explore a highly volatile system with $c=0.95$ and $r=2$.\\
In \fref{fig:HighlyVolatile} left panel, we show the richest (white), middle (grey) and poorest (black) third of the agents and the ACF (right panel) of the richest and poorest thirds. We see that the ACF for both have fat tails compared to the ACF of a random ensemble, reflecting the high degree of clustering seen in the left panel. The formation of large, macrocsopic neighbourhood areas for both the richest and the poorest third is clearly visible in this plot, whereas the middle third mostly forms a thin boundary layer between the two extremes instead of manifesting into large-scale neighbourhoods. Hence, the neighbourhood map shows a high polarization, reminiscent of the magnetic domains in ferromagnets. This superficial similarity should not be taken too seriously, though. The magnetic domains form because of energetic considerations of the macroscopic magnetic field, whereas our model and the Ising model are microscopic descriptions.

\onecolumngrid

\twocolumngrid

\end{document}